\documentclass[conference]{IEEEtran}
\IEEEoverridecommandlockouts
\usepackage[margin=0.7in]{geometry}

\usepackage{cite}
\usepackage{amsmath,amssymb,amsfonts}
\usepackage{algorithmic}
\usepackage{graphicx}
\usepackage{subfigure}
\usepackage{textcomp}
\usepackage{xcolor}
\usepackage{listings}
\usepackage[normalem]{ulem}
\usepackage{tabularx}

\begin{document}

\title{Fingerprinting of Machines in Critical Systems for Integrity Monitoring and Verification}

\author{Prakhar Paliwal, Arjun Sable, and Manjesh K. Hanawal, \\
        MLiONS Lab, Department of IEOR, IIT Bombay, Mumbai\\
}

\maketitle

\begin{abstract}
As cyber threats continue to evolve and diversify, it has become increasingly challenging to identify the root causes of security breaches that occur between periodic security assessments. This paper explores the fundamental importance of system fingerprinting as a proactive and effective approach to addressing this issue. By capturing a comprehensive host's fingerprint, including hardware-related details, file hashes, and kernel-level information, during periods of system cleanliness, a historical record is established. This historical record provides valuable insights into system changes and assists in understanding the factors contributing to a security breach. We develop a tool to capture and store these fingerprints securely, leveraging the advanced security features. Our approach presents a robust solution to address the constantly evolving cyber threat landscape, thereby safeguarding the integrity and security of critical systems. 
\end{abstract}


\section{Introduction}
In the modern era, marked by the growing impetus behind cyber threats, cybersecurity has risen to a position of paramount importance. With the abundance of sensitive data available online and the allure of financial gain through fraud and theft, cybercriminals continue to advance their tactics. To safeguard their assets, intellectual property, and personal information, governments, organizations, and individuals employ various methods, including vulnerability assessments of network architecture and endpoints, to fortify their defenses against malicious activities. However, the challenge arises when trying to discern what has transpired between the last assessment and a security breach. Uncovering the root cause of a breach is of utmost importance. This is where the significance of system fingerprinting emerges.

Fingerprinting in computer and network security is a fundamental process used to obtain comprehensive information about a system, encompassing its hardware, software, and configurations. This practice plays a pivotal role in various aspects of cybersecurity, including system identification, vulnerability assessment, network management, and digital forensics. By capturing a detailed snapshot of a system, its fingerprint establishes a valuable reference point for comparing the system's current state, enabling the detection of any alterations or modifications that may have transpired since the fingerprint was recorded.

The methodologies for capturing system fingerprints are essential tools in the arsenal of cybersecurity professionals, aiding them in maintaining the integrity and security of networked systems. This paper explores the key steps involved in the fingerprinting process for both Linux-based and Windows-based systems, shedding light on the techniques and tools used to gather critical information about a host's configuration and state. Understanding these methods is crucial for safeguarding against evolving cyber threats and addressing the ever-changing landscape of computer and network security. Our contributions are as follows:
\begin{itemize}
    \item We give a comprehensive list of system information that needs to be collected for system fingerprinting.
    \item We provide details on how to collect each of these system information and use it for forensic investigation. 
\end{itemize}

\section{Related work}

Kim and Spafford et al. (1993)\cite{kim1994tripwire} present Tripwire, an early and extensively utilized file system integrity checker specifically crafted for UNIX operating systems. Tripwire surveils system files and directories for unauthorized alterations by juxtaposing their present condition against a previously established signature database. The tool's adaptable, scalable, and portable characteristics render it appropriate for expansive network settings, providing system administrators with a mechanism to monitor modifications in system files and avert security violations. The implementation of cryptographic signatures within Tripwire guarantees both efficient and secure identification of file tampering.

In a parallel vein, Peddoju et al. (2020) \cite{peddoju2020file} concentrate on tackling the fundamental issues and obstacles encountered by File Integrity Monitoring (FIM) tools, accentuating the necessity of identifying unauthorized alterations in sensitive system files. Their investigation uncovers significant vulnerabilities in contemporary FIM tools, including delayed detection and deployment intricacies, which may leave systems susceptible to intrusions. To enhance the efficiency and efficacy of FIM, Peddoju et al. advocate for solutions that encompass real-time detection, covert operation, and tamper resistance, while simultaneously underscoring the drawbacks of host-based and guest-based tools in upholding file integrity.


\section{Information Collection for Fingerprinting}
\begin{table*}[h!]
\centering
\caption{System Fingerprinting Data Collection}
\begin{tabular}{|p{3.5cm}|p{10cm}|}
\hline
\textbf{Category} & \textbf{Description and Commands Used} \\ \hline
Installed Software & Inventory of software, Versions, and Paths. Collected using commands like \texttt{dpkg}, \texttt{rpm}, \texttt{wmic}, or \texttt{choco}. \\ \hline
Registry Keys & System-wide and user-specific configurations, startup settings. Queried using \texttt{reg query} (Windows) or equivalent tools. \\ \hline
Hardware Info & Motherboard details, BIOS data, PCIe slots, Network Adapters, and Sound Cards. Retrieved using \texttt{lspci}, \texttt{dmidecode}, or similar commands. \\ \hline
Network Config & DNS server details, IP routes, Static/DHCP settings. Collected using \texttt{ifconfig}, \texttt{ip a}, or \texttt{ip route}. \\ \hline
Host Details & Hostname, UUID, system type (laptop/desktop), boot ID, and OS name. Gathered using commands like \texttt{hostnamectl}. \\ \hline
Kernel/Firmware & Kernel version, CPU architecture, BIOS information. Extracted using \texttt{uname -r} or \texttt{dmidecode}. \\ \hline
Mounted Devices & Details of mounted volumes, storage devices, and motherboard configurations. Retrieved using \texttt{lsblk} or \texttt{df}. \\ \hline
Open Ports & Active ports, running services, and associated processes. Queried using \texttt{netstat} or \texttt{ss}. \\ \hline
Users/Groups & User accounts and Group memberships. Collected using \texttt{getent}, \texttt{id}, or \texttt{who}. \\ \hline
Scheduled Tasks & Task identifiers, execution timelines, and associated user accounts. Queried using \texttt{crontab -l} (Linux) or \texttt{schtasks} (Windows). \\ \hline
Container Info & Container metadata, including identifiers, runtime configurations, and operational statuses. Queried using \texttt{docker inspect}. \\ \hline
Extensions&  Browser extensions, IDE extensions (e.g., VSCode), and Python packages. Collected using browser APIs, \texttt{pip list}, or other relevant tools. \\ \hline
Secure Boot & Secure boot status, Bootloader integrity, and BIOS configurations. Queried using \texttt{efibootmgr} or similar tools.  \\ \hline
Time Zone & System time zone settings. Retrieved using \texttt{timedatectl} or \texttt{date} commands. \\ \hline
NTP Synced Status & Status of Network Time Protocol synchronization. Queried using \texttt{timedatectl} or \texttt{ntpq -p}. \\ \hline
Environment Variables & System and user-specific environment variables, including \texttt{PATH}, \texttt{HOME}, and application-specific configurations. Collected using \texttt{env} or \texttt{printenv}. \\ \hline

\end{tabular}
\label{tab:system_fingerprinting}
\end{table*}

A robust system fingerprint relies on the systematic compilation of diverse data that collectively enhance system security, detect vulnerabilities, and prevent unauthorized modifications. The subsequent elements are essential to this fingerprinting methodology, each providing distinct perspectives on the system’s configuration and integrity. In Table \ref{tab:system_fingerprinting}, we list the system information required for fingerprinting. Details of each information item is provided below: 
\subsection*{1. Installed Software and Packages}
A thorough catalog of the software and packages currently installed constitutes the fundamental basis of the system's software profile. This catalog records the names, versions and installation dates of applications which can also aid in vulnerability evaluations, particularly in the detection of obsolete softwares that are frequently targeted by malicious actors. Furthermore, documenting installation paths guarantees that software is located in anticipated directories, thereby reducing the likelihood of malware camouflaging itself by modifying these locations. Neglecting to uphold a precise inventory may allow adversaries to add or remove already installed softwares , establish persistence, or introduce backdoors.

For Linux, installed software can be listed using \texttt{dpkg-query -l} (Debian-based) or \texttt{rpm -qa} (Red Hat-based). On Windows, \texttt{wmic product get name,version} or \texttt{choco list --local-only} (Chocolatey) provides similar inventories.

\subsection*{2. Registry Keys and Subkeys}
The surveillance of registry keys and subkeys, particularly \texttt{HKEY\_LOCAL\_MACHINE} and \texttt{HKEY\_CURRENT\_USER}, is paramount for proficient system fingerprinting. The \texttt{HKEY\_LOCAL\_MACHINE} root key encapsulates system-wide configuration data, incorporating critical information regarding installed software, device drivers, and security settings, thereby rendering it indispensable for the identification of unauthorized alterations or malware that manipulate system-wide configurations. Essential subkeys such as \texttt{SOFTWARE}, \texttt{SYSTEM}, and \texttt{SECURITY} furnish insights into the operating system and hardware configurations. Conversely, \texttt{HKEY\_CURRENT\_USER} harbors user-specific configurations, encompassing application settings, preferences, and environmental variables, which are organized within subkeys such as \texttt{SOFTWARE}, \texttt{Control Panel}, and \texttt{Environment}. The monitoring of modifications in these domains is critical for the detection of user-level alterations, unauthorized software installations, or potential security threats that may compromise individual user sessions. 
\subsection*{3. Hardware Information}
Hardware profiling is essential for verifying system components' authenticity. Collecting motherboard and BIOS data is vital for firmware integrity, while monitoring connected devices helps identify unauthorized hardware. Proactive oversight of hardware changes is critical for preventing tampering and fostering trust in system components. Commands such as \texttt{Get-WmiObject} for Windows and \texttt{blkid} for Linux facilitate this data collection.
\subsection*{4. DNS and Network Configurations}
Network configurations delineate the communication framework underlying a system. The documentation of DNS servers ensures reliance on trusted resolvers, while the tracking of IP address allocations assists in identifying unexpected alterations indicative of spoofing or system compromise. Malicious alterations to network configurations can precipitate phishing attacks or unauthorized interception of network traffic, underscoring the necessity of vigilant monitoring of these elements.

For Linux, DNS servers can be listed using \texttt{cat /etc/resolv.conf} and IP addresses can be listed with \texttt{ip addr show} or \texttt{ifconfig}. On Windows, use \texttt{ipconfig /all} or \texttt{Get-DnsClientServerAddress} in PowerShell for DNS and IP information.

\subsection*{5. Host Information}
Host information, including hostname, UUID, system type, boot ID, and OS name, is vital for establishing a system's unique identity and configuration. This data is instrumental in differentiating devices on a network and allows for tailored security measures based on the type of system in use.

On Linux, the hostname can be retrieved using \texttt{hostname}, the UUID with \texttt{cat /sys/class/dmi/id/product\_uuid}, and system type via \texttt{uname -m}. The boot ID is available through \texttt{cat /proc/sys/kernel/random/boot\_id}, and the OS name can be obtained using \texttt{lsb\_release -a} or \texttt{cat /etc/os\_release}. On Windows, this information can be fetched with via command \texttt{systeminfo}

\subsection*{6. Kernel and Firmware Details}
The compilation of kernel and firmware information constitutes a fundamental component of system security. The documentation of the kernel version facilitates the proactive identification of vulnerabilities, while the monitoring of BIOS and firmware versions guarantees the legitimacy of updates and configurations. The exploitation of legacy kernels or firmware may lead to privilege escalation or system compromise, thereby underscoring the imperative for comprehensive monitoring.
For Linux, kernel details can be retrieved using \texttt{uname -r} and firmware information using \texttt{sudo dmidecode -t bios}. On Windows, \texttt{ver} displays the kernel version, while \texttt{wmic bios get name, version, serialnumber, releasedate} provides firmware details.

\subsection*{7. Mounted Devices and Baseboard Information}
Mounted devices signify a potential avenue for unauthorized access or data exfiltration. Fingerprinting incorporates specifics of mounted volumes and device identifiers to uncover unauthorized external storage or non-compliant devices. This proactive vigilance ensures compliance with data security protocols and alleviates risks associated with removable media or peripheral apparatus. Commands like \texttt{lsblk} or \texttt{df} commands, provides an overview of mounted devices in the system.
\subsection*{8. Open Ports and Services}
Open ports and services delineate the external-facing attack surface of the system. The collection of data regarding active ports and their associated processes facilitates the identification of unauthorized access points. The existence of unmonitored ports or unforeseen services amplifies the likelihood of exploitation, including data exfiltration or remote code execution. The fingerprinting of these components is crucial in minimizing the system's vulnerability to such attacks.

For Linux, open ports details can be retrived by \texttt{netstat -tuln}. On Windows, \texttt{netstat -an} lists the listening ports.

\subsection*{9. Users and Groups}
User and group information is crucial for understanding access control and privilege escalation risks. Monitoring user account identifiers ensures account validity and proper use of active accounts, while privilege levels confirm appropriate allocation of administrative rights. Systematic tracking of group memberships prevents unauthorized access elevation due to improper affiliations. Neglected accounts or excessive privileges increase vulnerability to data breaches and unauthorized system access. To fetch this information we can use commands like \texttt{net localgroup}, \texttt{net user} for windows based systems and \texttt{getent passwd}, \texttt{getent group} for linux based systems.
\subsection*{10. Scheduled Tasks}
Scheduled tasks are critical for the automation of operational routines; however, they concurrently serve as a potential vector for persistent threats posed by adversaries. The meticulous documentation of task identifiers, execution timelines, and associated user accounts is imperative to ensure that only sanctioned and intended tasks remain active. Malicious entities frequently exploit scheduled tasks to execute harmful payloads or facilitate data exfiltration. Conducting a comprehensive audit of
these tasks using commands like \texttt{schtasks /Query} for windows and \texttt{crontab} for linux is instrumental in fortifying the integrity of the system.
\subsection*{11. Container Information}
Containers represent lightweight, isolated environments that encapsulate applications alongside their dependencies. The assimilation of container metadata—featuring identifiers, status, runtime configurations, privilege, and image metadata (e.g., image ID and creation date)—within a system fingerprint is paramount for detecting unauthorized changes, as compromised containers can result in security incidents, impaired performance, and potential exploitation of the host system\cite{10041276}. In the Linux operating system, tools such as \texttt{docker inspect}, \texttt{podman inspect}, or \texttt{crictl} are capable of extracting both container and image metadata, whereas in the Windows environment, commands like docker ps and PowerShell modules such as DockerManagement provide analogous functionalities.
\subsection*{12. Extensions}
Extensions can enhance system capabilities, yet may create security risks if improperly managed. Systematic documentation of all extensions, including those for browsers and development environments like Visual Studio Code, enables the identification of unauthorized add-ons that could enable malicious activities. Continuous monitoring of these components is crucial for mitigating risks linked to sensitive data breaches and session hijacking through untrusted extensions.

\subsection*{13. Secure Boot}
Secure boot ensures the loading of only verified software during system initialization. A system fingerprint includes the status of secure boot and monitors for any modifications. Disabling secure boot or modifying the bootloader may allow malicious entities to install persistent malware, which can evade standard detection methods. To check whether secure boot is enabled or not, we can use command \texttt{mokutil --sb-state} and \texttt{efibootmgr -v} for linux based systems and \texttt{Get-SecureBootPolicy} for windows based systems.
\subsection*{14. Time Zone}
Time zone configuration and NTP synchronization are essential for precise system logs and event correlation in security analysis. They facilitate clock precision for time-sensitive operations and anomaly identification, encompassing log tampering.
Commands like \texttt{timedatectl status} can be used to check status for linux based systems and \texttt{w32tm /query /status} for windows based systems.

\begin{figure*}
\centering
\includegraphics[scale=.4]{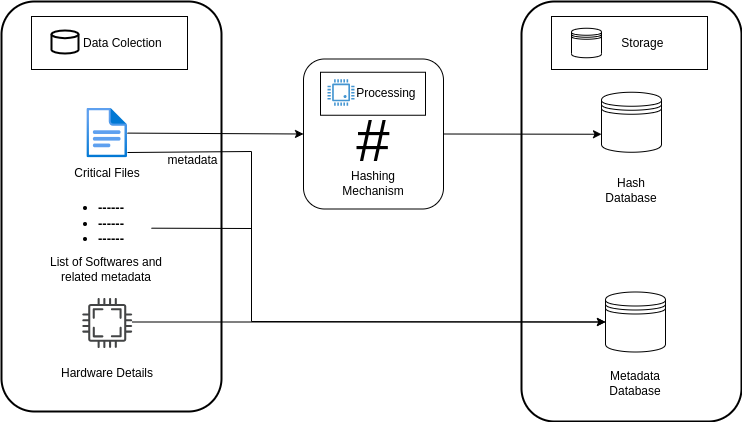}
  \caption{Proposed Steps to capture fingerprint  }
\end{figure*}

\section{Generating System Fingerprint}

When a system is sanitized based on Vulnerability and Assessment and Penetration testing reports, one can take the current snapshot of the system by taking its fingerprint. We refer to this as the baseline fingerprint, which can act as a reference for comparison with further fingerprints of the systems.  Based on the data acquired from both Linux and Windows operating systems, it is now feasible to construct a comprehensive system fingerprint baseline that encapsulates the fundamental attributes necessary for evaluating the integrity of these systems. This baseline constitutes a synthesis of diverse system metrics, including cryptographic hashes of critical files, registry data, and other relevant configurations documented in prior sections. We employ the SHA-256 cryptographic hash function to generate and preserve hashes for all essential directories and files. SHA-256 is widely regarded for its resilience against collision and pre-image attacks \cite{5076920}, making it an ideal choice for ensuring data integrity and detecting unauthorized alterations.

The subsequent phase involves hashing all critical directories containing sensitive system files and configurations. These directories are integral to the system's functionality, and any unauthorized changes can have significant security implications. Regular comparison of these hashes against the baseline enables the prompt detection of unauthorized modifications. Tables \ref{tab:windows} and \ref{tab:linux} in the appendix enumerate the critical directories for Linux and Windows systems, respectively, for which hash monitoring is being systematically conducted. This approach not only provides a temporal snapshot of the file contents but also strengthens the overall security architecture by ensuring the integrity of critical system files and directories.

Nevertheless, reliance exclusively on hashes may prove inadequate for comprehensive system integrity surveillance. To augment the efficacy of the fingerprint, it is imperative to monitor additional attributes, encompassing:

\begin{itemize}
    \item \textbf{File Size}: Variations in file size may signify corruption, tampering, or unauthorized modifications.
    \item \textbf{File Permissions}: Changes in file permissions could unveil potential attempts to alter access control settings, potentially exposing critical files to unauthorized individuals.
    \item \textbf{File Ownership}: Transformations in file ownership may indicate privilege escalation attempts or unauthorized user access.
    \item \textbf{Group Ownership}: Adjustments in group ownership could imply tampering with group-based access controls, potentially compromising system security.
\end{itemize}

These supplementary evaluations ensure a more holistic methodology towards system integrity, as articulated by Abdullah et al.\cite{abdullah2011file}, who underscore the necessity of implementing multiple integrity checks beyond mere hash comparisons to enhance detection accuracy and effectively mitigate security risks. By integrating file permissions, ownership, and size evaluations in conjunction with hash comparisons, we establish a fingerprint that exhibits greater resilience against covert attacks that aim to manipulate system properties without altering file content.

This baseline fingerprint now constitutes a thorough representation of a system's condition, encompassing its configurations, hardware specifications, installed applications, network configuration and other essential files, thereby serving as a reliable reference for monitoring and analytical purposes. Drift, which is characterized as deviations from this baseline, may signify potential security breaches or unauthorized alterations, including modified registry keys, unanticipated open ports, or the addition of hardware components. Through the continual comparison of the present system condition against the baseline, it becomes feasible to monitor for drift, facilitating the detection of discrepancies that could indicate tampering or other questionable activities. This methodology guarantees the swift identification of anomalies, thereby contributing to the preservation of system integrity and security.

Furthermore, it is crucial to ascertain the optimal frequency for capturing these fingerprints. Conducting integrity assessments at fixed intervals, while efficacious, may demand considerable system resources if executed too frequently, particularly on systems characterized by extensive directory structures and a substantial volume of files. Consequently, an adaptive scheduling strategy that modulates the frequency based on parameters such as file security level or system activity may prove advantageous. Therefor, Scheduling fingerprint monitoring according to system criticality optimizes security and resource utilization. This strategy reduces performance disruption while ensuring effective monitoring.


\section{Challenges}
While system fingerprinting proves indispensable in bolstering cybersecurity, several challenges merit careful consideration. Identifying crucial files for each node or host, customized to user and organizational needs, presents a formidable obstacle. Ensuring the tamper-proof nature of these identified files adds complexity, necessitating vigilant strategies. The dynamic nature of file systems introduces further intricacies, demanding the capability to distinguish genuine changes from potentially malicious ones. The proposed solution should address this by capturing fingerprints at regular intervals, reflecting the evolving state of systems. As explored in this research, the diverse methodologies for Linux-based and Windows-based systems contribute depth to the analysis, yet the challenge persists in implementing a unified approach that adapts to distinct operating environments. The amalgamation of robust file identification, tamper-proofing measures, and distinguishing between legitimate and unauthorized modifications underscores the complexity inherent in achieving effective system fingerprinting for cybersecurity purposes. Addressing these challenges is pivotal for ensuring the reliability, adaptability, and resilience of system fingerprinting practices in the ever-evolving landscape of cyber threats.

\section{Future Works}
In this work, we focused on fingerprinting Microsoft Windows and Linux operating systems. We can expand it to include Mac OS to broaden the applicability of our method. Also, 
Fingerprinting is one of the aspects of secure systems.  It could be used in conjunction with other cybersecurity tools to improve the security of the system.  One promising direction is to consolidate and optimize these methodologies by integrating them with existing Security Information and Event Management (SIEM) tools and Endpoint Detection and Response (EDR) systems \cite{9668567}. By doing so, it would enhance the capacity to monitor system integrity comprehensively and efficiently. Notably, this integrated module would be equipped to trigger alerts in response to any detected changes in the system, providing valuable insights for analysts to analyze and respond to potential security incidents. These future developments aim to elevate the effectiveness and versatility of system fingerprinting techniques in the ever-evolving landscape of cybersecurity.

\section{Conclusion}
In the face of escalating cyber threats, this research paper underscores the critical role of cybersecurity, exploring the fundamental process of system fingerprinting to safeguard sensitive data and assets. Vulnerability assessments serve as proactive measures, yet the challenge lies in discerning changes between assessments and security breaches. System fingerprinting, delved into meticulously for both Linux-based and Windows-based systems, emerges as a pivotal practice. This multifaceted approach, capturing detailed snapshots of a system's hardware, software, and configurations, provides a reference point for detecting alterations since the last assessment. The paper not only scrutinizes the methodologies and tools employed but also emphasizes the indispensable nature of system fingerprinting in cybersecurity. As an essential resource for cybersecurity professionals, this research contributes significantly to fortifying security measures and navigating the dynamic challenges of modern computing environments. In essence, the integration of robust system fingerprinting practices is key to resilient and adaptive cybersecurity strategies in our rapidly evolving digital landscape.
\label{sec:Conclusion}


\bibliographystyle{IEEEtran}
\bibliography{main}

\onecolumn
\section{Appendix}
\begin{table}[h!]
\centering
\caption{Windows Directories: Criticality and Monitoring Importance}
\begin{tabularx}{\textwidth}{|X|c|X|}
\hline
\textbf{File/Directory} & \textbf{Criticality} & \textbf{Importance of Monitoring} \\
\hline
C:\textbackslash & High & Root directory; any changes impact system operations and stability. \\
\hline
C:\textbackslash Windows & High & Operating system files; any changes can corrupt or compromise system functionality. \\
\hline
C:\textbackslash Program Files & Medium & Installed applications; modifications could introduce vulnerabilities or malware. \\
\hline
C:\textbackslash Users & Medium & User data and profiles; unauthorized access can expose sensitive information. \\
\hline
C:\textbackslash System32 & High & Core system binaries; tampering could disable key OS features or introduce malware. \\
\hline
C:\textbackslash Temp & Medium & Temporary files; excessive growth can indicate malicious activity or misconfigured applications. \\
\hline
C:\textbackslash AppData & High & Application data storage; unauthorized changes can affect program settings or compromise data. \\
\hline
C:\textbackslash Documents and Settings & Medium & Stores user profiles and configurations; modification could affect user environment. \\
\hline
C:\textbackslash ProgramData & High & Contains application data that is shared across users; tampering could affect the system’s behavior. \\
\hline
C:\textbackslash Recovery & High & System recovery files; alterations can prevent proper recovery during system failure. \\
\hline
C:\textbackslash \$Recycle.Bin & Medium & Deleted files storage; monitoring helps detect unauthorized file recovery attempts. \\
\hline
Windows Registry (HKEY\_LOCAL\_MACHINE, HKEY\_CURRENT\_USER) & High & Critical system settings; unauthorized changes can affect OS stability and security. \\
\hline

\hline
\end{tabularx}
\label{tab:windows}

\end{table}

\vspace{1cm} 

\begin{table}[h!]
\centering
\caption{Linux Directories: Criticality and Monitoring Importance}
\begin{tabularx}{\textwidth}{|X|c|X|}
\hline
\textbf{File/Directory} & \textbf{Criticality} & \textbf{Importance of Monitoring} \\
\hline
/ & High & Root directory; any changes impact overall system stability. \\
\hline
/bin & High & Holds essential binaries; unauthorized changes may compromise basic commands. \\
\hline
/boot & High & Contains boot files; modifications can disrupt startup or introduce malware. \\
\hline
/dev & Medium & Manages device files; unauthorized access could risk hardware control. \\
\hline
/etc & High & System-wide configurations; changes could alter system behavior or create backdoors. \\
\hline
/etc/cron & High & Manages scheduled tasks; modifications can introduce malicious tasks. \\
\hline
/etc/passwd & High & Contains user info; unauthorized edits can create or alter accounts. \\
\hline
/etc/shadow & High & Stores password hashes; tampering risks password security. \\
\hline
/lib, /lib64 & High & Shared libraries; changes may affect core binaries’ execution. \\
\hline
/root & High & Root user’s home; unauthorized changes can introduce malware. \\
\hline
/root/.ssh & High & SSH keys for root; compromise enables unauthorized root access. \\
\hline
/sbin & High & System binaries for root; changes can alter system functionalities. \\
\hline
/usr/lib & High & Libraries for user binaries; changes affect application behavior. \\
\hline
/usr/local/bin, /usr/local/lib, /usr/local/sbin & Medium & Stores locally installed binaries; changes could introduce unauthorized software. \\
\hline
Shell Binaries: ash, bash, dash, ksh, sh, tcsh & High & Core command interpreters; changes may compromise command input security. \\
\hline
Bash Config Files and Directories: /etc/bashrc, /etc/profile, .bashrc, .bash\_history, .bash\_login, .bash\_logout, .bash\_profile, .inputrc, .profile & Medium & Store shell settings; unauthorized edits may alter shell behavior or create persistence. \\
\hline
Common Log Files and Directories: /dev/log, /var/log, access\_log, auth.log, cron, dpkg.log, kern.log, last.log, mysql.log, mysqld.log, syslog, user.log, yum.log & High & System and activity logs; monitoring prevents tampering and ensures forensic accuracy. \\
\hline
Docker Binaries: /usr/local/bin/docker, /usr/local/bin/docker-compose, /usr/local/bin/notary & High & Critical for container management; changes could risk container security. \\
\hline
\end{tabularx}
\label{tab:linux}

\end{table}

\end{document}